\documentstyle[epsf]{europhys}

\def\stars{\bigskip\centerline{***}\medskip}

\newif\ifboo \boofalse

\newcommand{\bn}{{\mbox{{\boldmath $n$}}}}

\newcommand{\bx}{{\mbox{{\boldmath $x$}}}}
\newcommand{\by}{{\mbox{{\boldmath $y$}}}}
\newcommand{\bz}{{\mbox{{\boldmath $z$}}}}
\newcommand{\bone}{{\mbox{{\boldmath $1$}}}}
\newcommand{\ha  }{{1\over 2}}

\begin{document}
\euro{}{}{}{}
\Date{}
\shorttitle{Y. Kabashima {\em et al} STATISTICAL MECHANICS OF TYPICAL SET DECODING}

\title{Statistical mechanics of typical set decoding}
\author{
Yoshiyuki Kabashima$^{1}$\footnote{kaba@dis.titech.ac.jp},
Kazutaka Nakamura$^{1}$\footnote{knakamur@fe.dis.titech.ac.jp} and
Jort van Mourik$^{2}$\footnote{vanmourj@aston.ac.uk} }
\institute{$^{1}$Department of Computational Intelligence and Systems Science,
Tokyo Institute of Technology, Yokohama 2268502, Japan. \\
$^{2}$The Neural Computing Research Group, Aston
University, Birmingham B4 7ET, UK.\\}
\rec{}{}
\pacs{\Pacs{89}{90$+$n}{Other areas of general interest to physicists}
\Pacs{89}{70$+$c}{Information science}
\Pacs{05}{50$+$q}{Lattice theory and statistics; Ising problems}} 
\maketitle
\begin{abstract}
The performance of ``typical set (pairs) decoding'' 
for ensembles of Gallager's 
linear code is investigated using statistical physics. 
In this decoding, error happens when the information 
transmission is corrupted by an untypical noise 
or two or more typical sequences satisfy the parity 
check equation provided by the received codeword 
for which a typical noise is added. 
We show that the average error rate for the latter case 
over a given code ensemble can be tightly evaluated 
using the replica method, including the sensitivity to the message length. 
Our approach generally improves the existing 
analysis known in information theory community, which was 
reintroduced by MacKay (1999) and believed as most accurate to date. 
\end{abstract}

Triggered by active investigations on error correcting codes 
in both of information theory (IT) and statistical physics (SP)
communities 
\cite{MacKay,Richardson,Aji,us_EPL,us_PRL,Sourlas_nature,Sourlas_EPL,NishimoriWong}, 
there is a growing interest in the relationship between IT and SP. 
Since it turned out that the two frameworks 
that have different backgrounds have investigated 
similar subjects, it is quite natural to expect 
that standard techniques known in one framework
bring about remarkable developments in the other, and vice versa. 

The purpose of this Letter is to present such an example. 
More specifically, we will show that a method to 
evaluate the performance of error correcting codes 
established in IT community \cite{Aji,MacKay,Shannon_typical} 
can be generally improved 
by introducing the replica method. This serves as a direct answer 
to a question from IT researchers why the methods from physics 
always provide more optimistic evaluations than those known in IT literatures. 
In our formulation, the IT method is naturally linked to the existing 
SP analysis being parametrized by the number of replicas $\rho \ge 0$, 
which clearly explains how the IT and SP methods are related to each other. 

In a general scenario, the $N$ dimensional Boolean
message $\bx\in\{0,1\}^N$ is encoded to the $M(>N)$ dimensional
Boolean vector $\by^0$, and transmitted via a noisy channel, which is
taken here to be a Binary Symmetric Channel (BSC) characterized by
flip probability $p$ per bit; other transmission channels may also be
examined within a similar framework. At the other end of the channel,
the corrupted codeword is decoded utilizing the structured codeword
redundancy.

The error correcting code that we focus on here is Gallager's 
linear code \cite{Gallager}. This code was originally introduced by Gallgar
about forty years ago but was almost forgotten soon after the proposal 
due to the technological limitations in those days. 
However, since the recent rediscovery by MacKay and Neal \cite{MacKay}, 
this is now recognized as one of the best codes to date. 

A code of this type is characterized by a randomly generated $(M-N) \times
M$ Boolean sparse parity check matrix $H$, composed of $K$ and $C \
(\ge 3)$ non-zero (unit) elements per row and column,
respectively. Encoding the message vector $\bx$, is carried out using
the $M \times N$ generating matrix $G^T$, satisfying the condition
$HG^T\! = \!0$, where $\by^0\! = \!G^T \bx \ (\mbox{mod $2$})$.  The
$M$ bit codeword $\by^0$ is transmitted via a noisy channel, BSC in
the current analysis; the corrupted vector $\by\! = \!\by^0+\bn^0 \
(\mbox{mod $2$})$ is received at the other end, where
$\bn^0 \!\in\!\{0,1\}^M$ represents a noise vector with an independent
probability $p$ per bit of having a value 1. Decoding is carried out
by multiplying $\by$ by the parity check matrix $H$, to obtain the
syndrome vector $\bz\! = \!H \by \! = \!  H(G^T\bx+\bn^0)\! =
\!H\bn^0 \ (\mbox{mod $2$})$, and to find a solution to the parity 
check equation
\begin{eqnarray}
H \bn \! = \! \bz  \ (\mbox{mod $2$}) \ , 
\label{eq:parity_check}
\end{eqnarray}
for estimating the true noise vector $\bn^0$. One retrieves the original
message using the equation $G^T \bx \! = \! \by\! -\! \bn \ (\mbox{mod
$2$})$; $\bx$ becomes an estimate of the original message.

Several schemes can be employed for solving Eq.~(\ref{eq:parity_check}). 
In recent years, the maximum a posterior (MAP) and 
the maximizer of posterior marginal 
(MPM) decodings which correspond to zero and the Nishimori's 
temperatures, respectively, have been widely 
investigated \cite{Sourlas_EPL,Nishimori,Rujan,us_PRL}. 
However, we will here evaluate the performance of another scheme 
termed {\em typical set (pairs) decoding}, which was pioneered by 
Shannon \cite{Shannon_typical} and reintroduced by MacKay \cite{MacKay}
for analyzing the Gallager-type codes. 
Although this decoding method is slightly weaker to 
reduce the block or bit error rates, a rigorous analysis becomes 
easier than those for the above two decoding methods and investigation 
on it is now becoming popular in IT community \cite{Aji,Cover,MacKay}. 

In order to argue the typical set decoding, we first introduce 
the definition of {\em typical}. Due to the law of large number, 
a noise vector $\bn$ generated by the BSC satisfies a condition 
\begin{equation}
\left |\frac{1}{M}\sum_{l=1}^M n_l-p \right | \le \epsilon_M, 
\label{eq:typical}
\end{equation}
with a high probability for large $M$ and any sequence 
of positive number $\epsilon_M \sim {\cal O}(M^{-\gamma})$~
$(0< \gamma < 1/2)$. 
We define that a vector $\bn$ is 
classified as {\em typical} when this condition is satisfied. 
We also call the set of all typical vectors the {\em typical set}. 

Then, one can define the typical set decoding as a scheme to 
select a vector $\bn$ that belongs to the typical set
and satisfies Eq.~(\ref{eq:parity_check}), 
as an estimate of the true noise $\bn^0$. 
In the case that there are two or more typical vectors 
satisfy Eq.~(\ref{eq:parity_check}), it is decided that 
an error is automatically declared \cite{MacKay}. 
For this scheme, there can happen two types of decoding error; 
the first possibility, referred to the type I error here, 
takes place when the true noise $\bn^0$ is not typical, while 
the other one, termed the type II error, 
is declared when there are two or more typical vectors that satisfy 
Eq.~(\ref{eq:parity_check}) in spite that the true noise $\bn^0$ is typical. 
It can be shown that the probability for the type I error, 
$P_I$, vanishes in the limit 
$M \to \infty$. Therefore, we will here focus on the evaluation of 
the probability for the type II error, $P_{II}$.

To proceed further, it is convenient to employ the binary expression 
for bit sequences rather than Boolean one utilizing a
mapping $\{0,1,+\} \to \{+1,-1,\times \}$. 
This makes it possible to introduce the {\em error indicator} 
function that becomes one when error happens and zero, otherwise, as
\begin{eqnarray}
\Delta\left (\bn^0,H \right )=\lim_{\rho \to +0} {\cal V}_{NF}^\rho(\bn^0,H),
\label{eq:indicator}
\end{eqnarray}
where 
\begin{eqnarray} 
{\cal V}_{NF}(\bn^0,H) &\equiv& 
\mathop{\rm Tr}_{\bn \ne \bn^0}
\prod_{\mu=1}^{M-N} \delta\left (\prod_{l \in {\cal L}(\mu)} n^0_l,
\prod_{l \in {\cal L}(\mu)} n_l \right ) \delta \left (\sum_{l=1}^M n_l-M \tanh F \right ) 
\cr
&=&
\mathop{\rm Tr}_{\bn \ne \bone}
\prod_{\mu=1}^{M-N} \delta\left (1;
\prod_{l \in {\cal L}(\mu)} n_l \right ) 
\delta \left (\sum_{l=1}^M n^0_l n_l-M 
\tanh F \right ),  
\label{eq:Z_NF}
\end{eqnarray}
where we have introduced the gauge transform $n_l \to n^0_l n_l$ 
in the last form of Eq.~(\ref{eq:Z_NF}) for further convenience, and where
$\bone$ denotes the $M$ dimensional vector all the elements of which are $1$. 
Eq.~(\ref{eq:Z_NF}) denotes the number of vectors that differs from $\bn^0$ 
in the intersection of the typical set and the solution 
space of Eq. (\ref{eq:parity_check}). The field 
$F=(1/2)\ln \left [(1-p)/p \right]$
and ${\cal L}(\mu)$ represents the level of the channel noise 
and the set of indices that have non-zero elements 
in $\mu$ th row in the parity check matrix $H$, respectively.

From the definition, the probability of the type II error 
for a given matrix $H$ is given as $P_{II}(H)=\left \langle 
\Delta (\bn^0,H) \delta \left (\sum_{l=1}^M n^0_l -M \tanh F \right )
\right \rangle_{\bn^0}$, where 
$\left \langle \cdots \right \rangle_{\bn^0}=
\mathop{\rm Tr}_{\bn^0} \left (\cdots \right )\exp[\sum_{l=1}^M n^0_l ]/$
$\left (2 \cosh F\right )^M$. Since the parity check 
matrix $H$ is generated somewhat randomly, 
it is natural to evaluate the average of $P_{II}(H)$ over an ensemble of 
codes for given parameters $K$ and $C$ as a performance measure for 
the code ensemble. Employing Eq.~(\ref{eq:indicator}), 
the average is given as
$\overline{P_{II}}  =  \lim_{\rho \to +0} \exp \left [-M 
{\cal E}(\rho) \right ]
$, 
where 
\begin{eqnarray}
{\cal E}(\rho) \equiv - \frac{1}{M} \ln \left \langle 
\left \langle 
{\cal V}^\rho _{NF}(\bn^0,H) \delta \left (\sum_{l=1}^M n^0_l -M \tanh F \right )
\right \rangle_{\bn^0} \right \rangle_H, 
\label{eq:E_rho}
\end{eqnarray}
for large $M$. Here, $\left \langle \cdots \right \rangle_H$ 
represents an average over the uniform 
distribution of the parity check matrix for 
a given choice of parameters $K$ and $C$. 

Before proceeding any further, it is worthy of mentioning 
general properties of the exponent ${\cal E}(\rho)$. 
First, $\overline{P_{II}}$ is expected to vanish in the 
limit $M \to \infty$ for a sufficiently small noise $p$. 
The highest noise level $p_c$ for this is termed 
the {\em error threshold} \cite{Aji}. 
This happens when ${\cal E}(0)=\lim_{\rho \to +0} {\cal E}(\rho)>0$. 
The value of ${\cal E}(0)>0$ represents the sensitivity 
of $\overline{P_{II}}$ to the message length and 
serves as a performance measure of the code ensemble when $M$ is finite. 
Next, since ${\cal V}_{NF}(\bn^0,H)=0,1,2,\ldots$, 
${\cal V}_{NF}^\rho(\bn^0,H)$ increases with respect 
to $\rho$, which implies the exponent ${\cal E}(\rho)$ becomes a decreasing 
function of $\rho \ge 0$. 
This is linked to an inequality 
\begin{eqnarray}
\frac{\partial {\cal E}(\rho)}{\partial \rho}
=-\frac{1}{M}\frac{\left \langle 
\left \langle 
{\cal S}_{NF}(\bn^0,H) {\cal V}^\rho _{NF}(\bn^0,H) 
\delta \left (\sum_{l=1}^M n^0_l -M \tanh F \right )
\right \rangle_{\bn^0} \right \rangle_H}
{
\left \langle 
\left \langle 
{\cal V}^\rho _{NF}(\bn^0,H) 
\delta \left (\sum_{l=1}^M n^0_l -M \tanh F \right )
\right \rangle_{\bn^0} \right \rangle_H}
< 0, 
\label{eq:entropy}
\end{eqnarray}
where ${\cal S}_{NF}(\bn^0,H)=\ln {\cal V}_{NF}(\bn^0,H)$
is the entropy representing the number of wrong solutions 
for Eq.~(\ref{eq:parity_check}) belonging to the typical set. 
One can also show that $\partial^2 {\cal E}(\rho)/\partial \rho^2 < 0$, 
which implies ${\cal E}(\rho)$ is a convex function of $\rho$. 

We are now ready to connect the current argument to the existing 
analysis of the typical set decoding \cite{Shannon_typical,MacKay,Aji}. 
Since ${\cal E}(0) \ge {\cal E}(1)$, one can obtain a {\em lower} bound of 
$p_c$ from the condition ${\cal E}(1)\!=\!0$. 
For $\rho\!=\!1$ in Eq.~(\ref{eq:E_rho}), it is convenient to insert 
an identity $1\!=\!\int d\omega \ \delta 
\left (\sum_{l=1}^M n_l -M \omega \right )$
in the final form of Eq.~(\ref{eq:Z_NF}). 
Then, for a sequence $\bn$ that satisfies $(1/M)\sum_{l=1}^M n_l=\omega$, 
one obtains
$
\left \langle \mathop{\rm Tr}_{\bn} 
\delta \left (\sum_{l=1}^M n^0_l n_l -M \tanh F 
\right) \delta \left ( \sum_{l=1}^M n^0_l-M \tanh F \right ) \right
\rangle_{\bn^0} $
$ \sim \exp \left [-M {\cal K}(\omega,F) \right ]$, 
where
${\cal K}(\omega,F)
\!=\! \left (\frac{1\!+\!\omega}{2} \right )
 H\left (\frac{2 \tanh F}{1\!+\!\omega} \right )
\!+\!\left (\frac{1\!-\!\omega}{2} \right )\ln 2 -\! H(\tanh F)
$
and $H(x)\!=\!-\frac{(1+x)}{2} 
\ln \frac{(1\!+\!x)}{2}-\frac{(1\!-\!x)}{2} \ln \frac{(1-x)}{2}$. 
Further, the remaining average required in Eq.~(\ref{eq:E_rho}) is 
evaluated as 
$
\left \langle \mathop{\rm Tr}_{\bn} \delta \left (\sum_{l=1}
n_l \right . \right . $
$ \left . \left . 
-M \omega 
\right )
\right .
$
$
\left . 
\prod_{\mu=1}^{M-N} 
\delta \left (1;\prod_{l \in {\cal L} (\mu ) } n_l \right ) \right \rangle_{H}
\sim \exp\left [ M {\cal R}\left (\omega \right ) \right ]. 
$
The exponent 
$
{\cal R}\left (\omega \right ) 
$
is termed the {\em weight enumerator} \cite{Aji,MacKay}.
This provides an averaged distribution of the distances between 
the true noise $\bn^0$ and 
other vectors that satisfy 
Eq.~(\ref{eq:parity_check}) 
in the current context\footnote{
The weight enumerator is usually introduced for 
the distance between codewords \cite{Aji,MacKay,McEliece}. 
However, since $\by^0-\by^1=\bn^0-\bn^1 \ \mbox{(mod $2$)}$ 
holds for two sets of Boolean vectors $(\by^0,\bn^0)$ and 
$(\by^1,\bn^1)$ that satisfy $\by=\by^0+\bn^0=\by^1+\bn^1 \ \mbox{(mod $2$)}$, 
the distance between the noise vectors $\bn^0$ and $\bn^1$ is 
identical to that for the codewords $\by^0$ and $\by^1$. 
}, 
and plays an important role for evaluating a performance 
of codes in conventional coding theory \cite{McEliece}. 
From the above, one obtains 
${\cal E}(1)=\mathop{\rm Ext}_{\omega \ne 1}\left \{
{\cal K}(\omega,F)-{\cal R}(\omega) \right \}$, where 
$\mathop{\rm Ext}_{\{\cdots\}}$ denotes an extremization.  
This corresponds to Eq.~(4.7) in \cite{Aji}. 

However, it should be emphasized here that one can evaluate ${\cal E}(1)$ 
without introducing the weight variable $\omega$. Moreover, 
it is evident that the tightest estimate (exact value) of $p_c$ 
can be obtained by evaluating ${\cal E}(0)=\lim_{\rho \to +0}{\cal
E}(\rho)$. 
This can be carried out by the replica method, 
which gives rise to a set of order parameters
$q_{\alpha, \beta, \ldots,\gamma}=(1/M) \sum_{l=1}^M Z_l 
n_l^\alpha n_l^\beta \ldots 
n_l^\gamma$, where $\alpha, \beta, \ldots$ represent replica indices and the 
variable $Z_{l=1,\ldots,M}$ comes from enforcing the 
restriction $C$ connections 
per index as in \cite{us_PRL}. 

Further calculation requires a certain ansatz about 
the symmetry of the order parameters. 
As a first approximation we assume replica symmetry (RS) in 
the following order parameters and their conjugate variables
\begin{eqnarray}
q_{\alpha, \beta, \ldots,\gamma}=q \int
dx \ \pi (x) \ x^l, \quad  
\widehat{q}_{\alpha, \beta, \ldots,\gamma}=\widehat{q} \int
d\widehat{x} \ \widehat{\pi} \ (\widehat{x}) \widehat{x}^l, 
\label{eq:order_parameters}
\end{eqnarray}
where $l$ denotes the number of replica indices, $q$ and $\widehat{q}$ 
are normalization 
variables for defining 
$\pi(\cdot)$ and $\widehat{\pi}(\cdot)$ as distributions. 
Unspecified integrals are carried out over the interval
$[-1,1]$. 

Originally, the summation $\mathop{\rm Tr}_{\bn \ne \bone}(\cdot)$ excludes 
the case $\bn =\bone$; but one can show that for large $M$ limit, this becomes
identical to the full summation in the non-ferromagnetic phase, where 
$\pi(x) \ne \delta(x-1)$ and 
$\widehat{\pi}(x) \ne \delta(\widehat{x}-1)$. 
In addition, we employ Morita's scheme \cite{Morita} which in this case
converts 
the restricted annealed average with respect to $\bn^0$ to a quenched 
one:
\begin{eqnarray}
\frac{1}{M} \ln \left \langle (\cdots) \times \delta 
\left ( \sum_{l=1}^M n^0_l -M \tanh F \right ) \right \rangle_{\bn^0}
=\frac{1}{M} \left \langle \ln (\cdots ) \right \rangle_{\bn^0}, 
\label{eq:Morita}
\end{eqnarray}
to simplify the calculation of the average over $\bn^0$ in
Eq.~(\ref{eq:E_rho}) considerably, and obtain
\begin{eqnarray}
{\cal E}(\rho) \! &=& \!
\mathop{\mbox{\rm Ext}^*}_{ \{ q,\widehat{q}, 
\pi(\cdot),\widehat{\pi}(\cdot), G \} }
\left \{
-\frac{C \ q^K}{K} \int \prod_{i\! = \!1}^K d x_i \pi(x_i) 
\left 
(\frac{ 1+ \prod_{i\! = \!1}^K x_i}{2} \right )^\rho 
\right . \cr 
&-& \left \langle \ln \left [
\int \prod_{\mu\! = \!1}^C d \widehat{x}_{\mu} \ \widehat{\pi} (\widehat{x}_{\mu})
\left (\mathop{\rm Tr}_{n=\pm 1}
 e^{ G n^0 n } 
\prod_{\mu\! = \!1}^C \left ( \frac{1+\widehat{x}_\mu \  n }{2} \right ) 
\right )^\rho \right ]  \right \rangle_{n^0}\cr
&-& \left . C \ln \widehat{q}\!+\! 
C q \widehat{q} \int dx \ d \widehat{x} \ \pi(x) \ \widehat{\pi}(\widehat{x}) 
\left (\frac{ 1+ x \widehat{x}}{2} \right )^\rho 
+\left ( \frac{C}{K}-C \right ) +\rho\  G \tanh F
\right \}, 
\label{eq:E_replica}
\end{eqnarray}
where $\left \langle (\cdots)\right \rangle_{n^0}=\mathop{\rm
Tr}_{n^0=\pm 1} (\cdots)$ and $\mathop{\rm Ext}_{\{\cdots\}}^*$ denotes
the functional extremization excluding the possibility of 
$\pi(x)=\delta(x-1)$ and 
$\widehat{\pi}(\widehat{x})=\delta(\widehat{x}-1)$ as
is introduced in \cite{Reliability}. 

Two analytical solutions of $\pi(x)$ and 
$\widehat{\pi}(\widehat{x})$ can be obtained in the limit
$K,C \to \infty$, keeping the code rate $R=N/M=1-C/K$ finite: 
\begin{enumerate}
\item
$\pi(x)=\ha[(1+\tanh F)\delta(x-\tanh F)+(1-\tanh F) \delta(x+\tanh F)]$,~~~
$\widehat{\pi}(\widehat{x})=\delta(\widehat{x})$
\item
$\pi(x)=\ha[\delta(x-1)+\delta(x+1)]$,~~~
$\widehat{\pi}(\widehat{x})=\ha [\delta(\widehat{x}-1)+
\delta(\widehat{x}+1)]$. 
\end{enumerate}
One can show that both of these are locally stable against 
perturbations to the RS solutions providing 
${\cal E}(\rho)=\rho~\left[ H(\tanh F)\!-\!(1\!-\!R)\ln 2 \right ]$
and ${\cal E}(\rho)=H(\tanh F)\!-\!(1\!-\!R)\ln 2$, respectively. 
Selecting the relevant branch that has the lower 
exponent for $\rho \ge 1$ and 
taking the limit $\rho \to 0 $ \cite{Hemmen_Palmer}, 
one obtains the exponent as
\begin{eqnarray}
{\cal E}(0)=\lim_{\rho \to +0} {\cal E}(\rho)=
\left \{
\begin{array}{ll}
(R_c-R)\ln2 , & \mbox{$R<R_c$,} \cr
0, & \mbox{$R>R_c$,} 
\end{array}
\right .
\end{eqnarray}
where $R_c=1+p\log_2 p + (1-p) \log_2 (1-p)$ corresponds to 
Shannon's limit \cite{Shannon}. 

Note that in the vicinity of $R \sim R_c$, this exponent
exceeds the {\rm upper} bound of possible reliability function 
that represents the vanishing rate of the decoding error probability for the
best code \cite{McEliece_Omura,Reliability}. 
However, this does not imply any contradiction because 
the current analysis is just for $\overline{P_{II}}$ while 
the convergence rate of $P_I$ is slower than that of the reliability function. 

For finite $K$ and $C$, one can obtain ${\cal E}(\rho)$ 
via numerical methods. Similar to the case of $K,C \to \infty$, 
there generally appear two branches of solutions:
\begin{enumerate}
\item 
continuous distributions for
$\pi(x)$ and $\widehat{\pi}(\widehat{x})$, for which $\lim_{\rho
\to +0} {\cal E}(\rho)=0$. 
\item
$\rho$ independent frozen distributions 
$\pi(x)=\ha [(1+b)~\delta(x-1)+(1-b)~\delta(x+1)]$, \\
$\widehat{\pi}(\widehat{x})=\ha [(1+\widehat{b})~
\delta(\widehat{x}-1)+(1-\widehat{b})~\delta(\widehat{x}+1)]$.
\end{enumerate}
The parameters $b$ and $\widehat{b}$ are determined from the extremization 
problem (\ref{eq:E_replica}) by setting $\rho=1$, the functional extremization
with respect to $\pi(\cdot)$ and $\widehat{\pi}(\cdot)$ is 
then reduced to that for the first moments 
$b=\int dx \pi(x)$ and $\widehat{b}=\int d \widehat{x} 
\widehat{\pi}(\widehat{x})$. 
The exponent of this branch is completely frozen to that for $\rho=1$
as ${\cal E}(\rho)={\cal E}(1)$ for $\forall{\rho} > 0$. 
Although the distributions of the two branches look quite different,
their exponents coincide at $\rho=1$ in any situation. 

\begin{figure}[t]
\setlength{\unitlength}{1mm}
\begin{picture}(136,31)
\epsfxsize=136 mm
\put(0,0){
\epsfbox{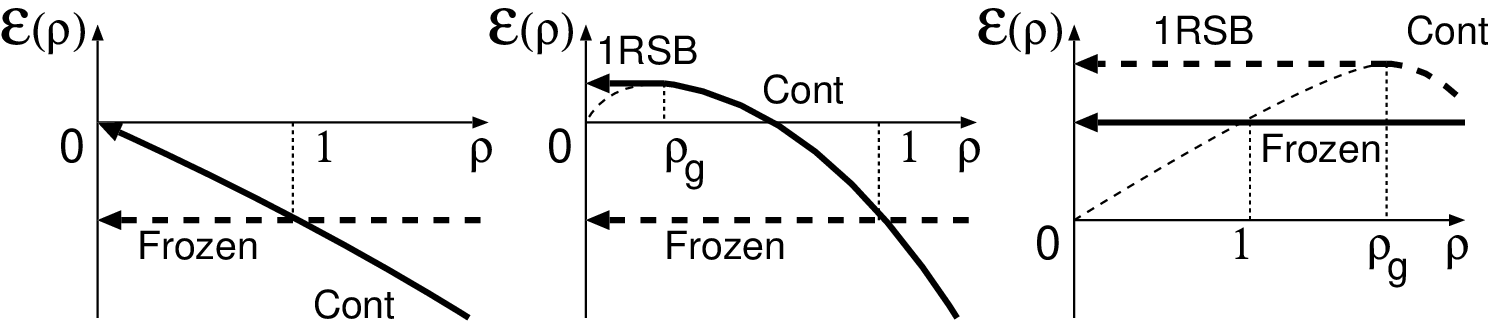}}
\put(8,31){(a)}
\put(52.5,31){(b)}
\put(97,31){(c)}
\end{picture}
\caption{Appropriate limits for 
$\lim_{\rho \to +0} {\cal E}(\rho)$ in the case of finite $K$ and $C$. 
The solution that has the lower exponent for $\rho \ge 1$ 
should be selected as the relevant branch \cite{Hemmen_Palmer}, 
which is drawn as a thick curve or line in each case. 
For $p \ge p_c$ (a), the continuous solution is relevant while the 
1(frozen)RSB solution which emerges
from this solution at $\rho=\rho_g$ 
provides an appropriate exponent ${\cal E}(\rho_g)$
for $p_b \le p < p_c$ (b). For $0 < p < p_b$ (c), 
the frozen (RS) solution is relevant. 
In the limit $K,\ C \to \infty$, the situation (b) does not appear. 
}
\label{fig:e_rho}
\setlength{\unitlength}{1mm}
\begin{picture}(140,51)
\epsfysize=48 mm
\put(-10,0){\epsfbox{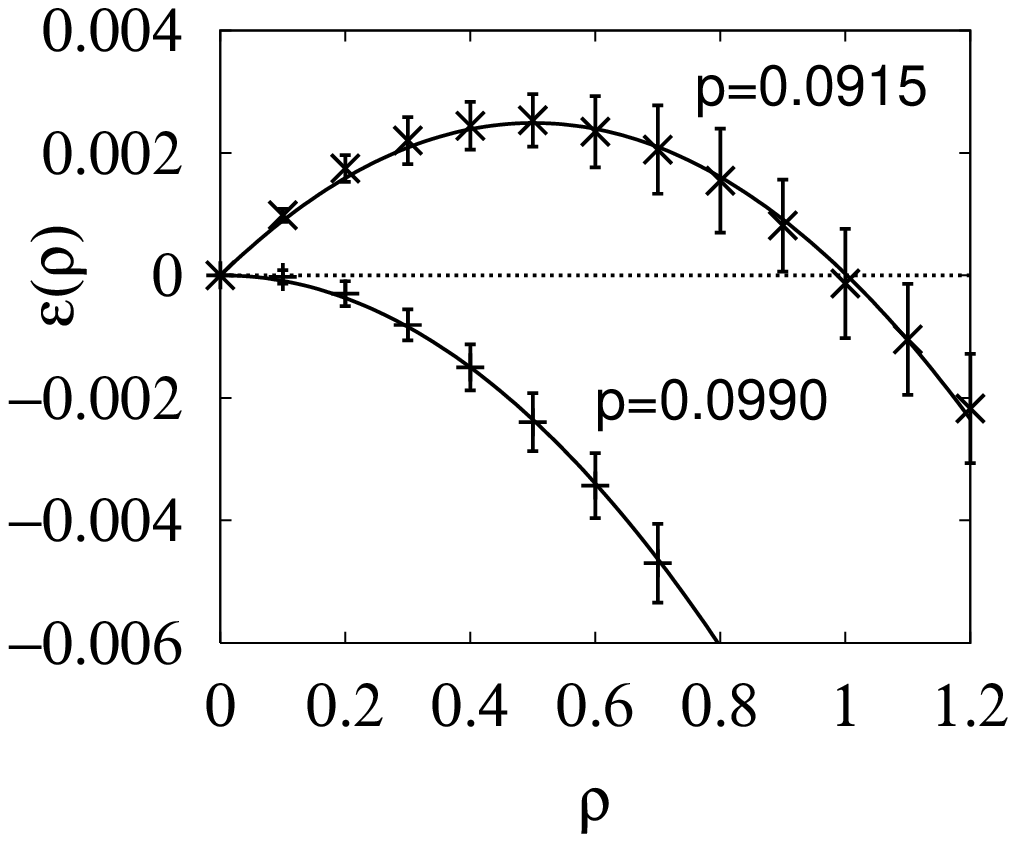}}
\put(10,48){(a)}
\put(70,48){(b)}
\put(60,30){
\begin{tabular}{c|cccc}
$(K,C)$ & $(6,3)$ & $(5,3)$ & $(6,4)$ & $(4,3)$ \\
\hline
Code rate  & $1/2$ & $2/5 $ & $1/3 $ & $1/4 $ \\
\hline
IT & 0.0915 & 0.129 & 0.170 & 0.205 \\
\hline
Current Method & 0.0990 & 0.136 & 0.173 & 0.209 \\
\hline
Shannon's limit & 0.109 & 0.145 & 0.174 & 0.214 \\
\hline
\end{tabular}
}
\end{picture}
\caption{(a): Numerically computed ${\cal E}(\rho)$ of the 
continuous branch for $p=0.0915, \ 0.0990$ for $K=6$ and $C=3$ ($R=1/2$). 
Symbols and error bars are obtained from $50$ numerical solutions. 
Curves are computed via a quadratic fit. For $p=0.0915$, ${\cal E}(\rho)$ 
is maximized to a positive value ${\cal E}(\rho_g) 
\simeq 2.5 \times 10^{-3}$ for 
$\rho_g \simeq 0.5$ while it vanishes 
at $\rho \simeq 1$ as is suggested in the IT literature \protect{\cite{Aji}}. 
On the other hand, for $p=0.0990$, our predicted threshold, 
it is maximized to zero at $\rho \simeq 0$, which implies
that this is the correct threshold. 
(b): Comparison of the estimates of $p_c$ between the IT 
and the current methods 
is summarized in a table. The estimates for the IT method are taken 
from \cite{Aji}. The numerical precision is up to the last 
digit for the current method. Shannon's limit denotes the highest 
possible $p_c$ for a given code rate. 
}
\label{fig:numerical}
\end{figure}
Note that the frozen branch corresponds to the conventional IT analysis
\cite{Aji,MacKay}, and would provide the exact estimate in absence of
other solutions. However, in order to take an appropriate limit
$\lim_{\rho \to +0}{\cal E}(\rho)$, one has to select the dominant branch for 
$\rho \ge 1$ \cite{Hemmen_Palmer} among the existing solutions, and the frozen
branch does not necessarily provide the correct exponent for $\rho \to +0$.
Actually, the scenario suggested by our analysis supports this statement
(Fig.~\ref{fig:e_rho}). 

When the channel noise $p$ is sufficiently high (Fig.~\ref{fig:e_rho} (a)), the
exponent for the continuous branch is monotonically decreasing with respect to 
$\rho$ which implies this is the dominant branch for $\rho \ge 1$. 
This provides $\lim_{\rho \to +0}{\cal E}(\rho)=0$. 
However, for lower $p$, ${\cal E}(\rho)$ of the continuous 
branch is maximized to a positive value for a certain parameter
$\rho_g$(Fig.~\ref{fig:e_rho} (b)). 
In this situation, the solution for $0 < \rho < \rho_g$ is physically
wrong because inequality (\ref{eq:entropy}) 
does not hold. The frozen replica symmetry 
breaking (RSB) solution \cite{Gross} 
(a one step RSB ansatz under the constraint $(1/M)\bn^a \cdot \bn^b=1$ 
for replica indices $a$ and $b$ in the same subgroup) is a suitable 
scheme for obtaining a consistent solution. Employing this 1RSB solution, 
one finds ${\cal E}(\rho)={\cal E}(\rho_g)$ for $0 < \rho < \rho_g$, 
which implies $\lim_{\rho \to +0}{\cal E}(\rho)={\cal E}(\rho_g) > 0$ 
indicating a vanishing behaviour
$\overline{P_{II}} \sim \exp\left [-M {\cal E}(\rho_g)\right ]$. 
These imply that the critical condition 
determining the error threshold $p_c$ is given 
by $\left. \partial {\cal E}(\rho )/ 
\partial \rho \right |_{\rho \to +0} =0 $, 
being computed for the continuous solution. 
Employing the gauge transform \cite{Nishimori}, one can show that 
the variational parameter $G$ in Eq.~(\ref{eq:E_replica}) enforcing 
$\sum_{l=1}^M n_l^0 n_l = M \tanh F$ coincides with $F$ in this 
limit. Then, the critical condition is summarized as
\begin{eqnarray}
F \tanh F -\frac{1}{M} \left \langle \left \langle 
\ln \left [
\mathop{\rm Tr}_{\bn \ne \bone} \prod_{\mu=1}^{M-N}
\delta \left (1; \prod_{l \in {\cal L}(\mu) }n_l \right )
e^{F \sum_{l=1}^M n^0_l n_l}
\right ] \right \rangle_H \right \rangle_{n^0}=0, 
\label{eq:transition}
\end{eqnarray}
which is identical to what has been obtained for the phase boundary
of the ferro-paramagnetic transition along the Nishimori's 
temperature predicted by the existing replica 
analysis \cite{us_PRL,Reliability}. 

As $p$ is reduced further, the position of the maximum $\rho_g$
moves to the right and exceeds $\rho=1$
at another critical noise rate $p_b$. 
This implies that below $p_b$ the limit $\rho \to +0$ is governed 
by the frozen (RS) solution that is identical to what is given by 
the conventional IT analysis (Fig.~\ref{fig:e_rho} (c)). 
However, this situation is realized only sufficiently below from the
threshold and the solution is of no use for direct evaluation of $p_c$ 
although it provides a lower bound. 

Finally, we examined the case of $K=6$ and $C=3$ to demonstrate the 
accuracy of the estimated threshold. 
We numerically evaluated ${\cal E}(\rho)$ of the 
continuous branch for $p=0.0915$, a recent highly 
accurate estimate of the error threshold for this parameter
choice \cite{Aji}
and for $p=0.0990$, which is the threshold predicted by the replica
method \cite{Nakamura,Reliability}. 
The numerical results are obtained by approximating $\pi(\cdot)$ and
$\widehat{\pi}(\cdot)$ using $10^6$ dimensional vectors 
and iterating the saddle point equations until convergence. 
The obtained results are shown in Fig.~\ref{fig:numerical} (a);
it indicates $\mathop{\rm max}_{\rho } {\cal E}(\rho)
\simeq 2.5 \times 10^{-3}$ 
for $p=0.0915$ while ${\cal E}(\rho)$ is maximized 
(to zero) at $\rho \simeq 0$ for $p=0.0990$, 
suggesting a tighter estimate for the error threshold than those 
reported so far. Comparison in other parameter choices is 
also summarized in Fig. \ref{fig:numerical} (b). 

In summary, we have investigated the performance of the 
typical set decoding for ensembles of Gallager's 
codes. We have shown that the direct evaluation of the average 
type II error probability over the ensemble becomes possible employing the 
replica method. The link to the existing IT 
analysis which is based on the weight enumerator 
is also clarified. Although the weight enumerator 
does not play a crucial role for determination of the error 
threshold in the current analysis, it still remains a key 
factor for the error rate in low $R$ regions. 
Analysis of it from a view point of 
statistical physics is under way \cite{Jort}. 

\stars

We acknowledge support from the Grant-in-Aid, 
the Japan-Anglo Collaboration Programme of the JSPS (YK), 
EPSRC (GR/N00562) and The Royal Society (JVM).
David Saad is acknowledged for useful comments and discussions.


\begin{thebibliography}{99} 
%
%
%
%
\bibitem{Aji} S.~Aji, H.~Jin, A.~Khandekar, D.J.C.~MacKay and R.J.~McEliece, 
BSC Thresholds for Code Ensembles Based on ``Typical Pairs'' Decoding, 
preprint, (1999). 
%
\bibitem{Cover} T.M.~Cover and J.A.~Thomas, {\em Elements of 
Information Theory}, 
Wiley (New York), (1991). 
%
\bibitem{Gallager} R.G.~Gallager, {\em IRE Trans.~Info.~Theory}, {\bf
IT-8}, 21 (1962).
%
\bibitem{Gross} D.J.~Gross and M.~M\'{e}zard, {\em Nucl. Phys. }, {\bf
B240}, 431 (1984).
%
\bibitem{Hemmen_Palmer} J.L.~van~Hemmen and R.G.~Palmer, 
{\em J. Phys. A: Math. and Gen.}, {\bf 12}, 563 (1979).
%
\bibitem{us_EPL} Y.~Kabashima and D.~Saad, {\em Europhys.~Lett.}, {\bf
44}, 668 (1998); {\bf 45}, 97 (1999).
%
\bibitem{us_PRL} Y.~Kabashima, T.~Murayama and D.~Saad, {\em Phys.Rev.Lett.}, 
{\bf 84}, 1355 (2000); T.~Murayama, Y.~Kabashima, D.~Saad and R.~Vicente, 
{\em Phys. Rev. E}, {\bf 62}, 1577 (2000).
%
\bibitem{Reliability} Y.~Kabashima, N.~Sazuka, K.~Nakamura and 
D.~Saad, cond-mat/0010173 (2000). 
%
\bibitem{MacKay} D.J.C.~MacKay, {\em IEEE Trans. on Info. Theor}, {\bf 45}, 
399 (1999); D.J.C.~MacKay and R.M.~Neal, {\em Electronic Lett.}, 
{\bf 33}, 457 (1997).
%
\bibitem{McEliece} R.J.~McEliece, {\em The Theory of Information and Coding}, 
Addison-Wesley (Reading, MA), (1977). 
%
\bibitem{McEliece_Omura} R.J.~McEliece and J.~Omura, 
{\em IEEE Trans. on Infor. Theor}, {\bf 23}, 611 (1977). 
%
\bibitem{Morita} T.~Morita, {\em Math. Phys.} {\bf 5}, 1401, (1964);
R.~K\"uhn, {\em Z. Phys. B} {\bf 100}, 231 (1996).
%
\bibitem{Jort} J.~van~Mourik, D.~Saad and Y.~Kabashima, preprint (2001). 
%
\bibitem{Nakamura} K.~Nakamura, Y.~Kabashima and D.~Saad, 
cond-mat/0010073 (2000).  
%
\bibitem{Nishimori} H.~Nishimori, {\em J.~Phys.~Soc.~of Japan},
{\bf 62}, 2973 (1993).  
%
\bibitem{NishimoriWong} H.~Nishimori and K.Y.M.~Wong, 
{\em Phys. Rev. E}, {\bf 60}, 132 (1999).  
%
\bibitem{Richardson} T.~Richardson, A.~Shokrollahi and R.~Urbanke, 
Design of provably good low-density parity check codes, preprint (1999)
%
\bibitem{Rujan} P.~Ruj\'{a}n, {\em Phys.~Rev.~Lett.}, {\bf 70},
2968 (1993).
%
\bibitem{Shannon} C.E.~Shannon, {\em Bell Sys.~Tech.~J.}, {\bf 27},
379 (1948); {\bf 27}, 623 (1948).
%
\bibitem{Shannon_typical} C.E.~Shannon, {\em The Mathematical Theory of 
Information}, University of Illinois Press (Urbana, IL), (1949);
reprinted (1998). 
%
\bibitem{Sourlas_nature} N.~Sourlas, {\em Nature}, {\bf 339}, 693
(1989).  
%
\bibitem{Sourlas_EPL} N.~Sourlas, {\em Euro.Phys.Lett.}, {\bf 25}, 159
(1994).
\end{thebibliography}
\end{document}